\documentclass[letterpaper]{article}
\usepackage{multicol}
\usepackage{fullpage}
\usepackage{amsmath}
\usepackage{amsfonts}

\title{Cryptographic Authentication of Navigation Protocols}
\author{
        Sam Bretheim \\
        bret0045@umn.edu
        }
\date{}

\nonstopmode
\begin{document}

\maketitle

\begin{multicols}{2}

\begin{abstract}
We examine the security of existing radio navigation protocols and attempt to define secure, scalable replacements.
\end{abstract}

\section{Motivation}

Resourceful and well-funded groups are clearly interested in using aircraft as projectile weapons. Increased airport and cockpit security can significantly increase the difficulty of bombings and hijackings, but some attacks can be executed without ever touching their targets: It is possible to lead an aircraft astray by transmitting noise or false information to its communication and navigation systems, potentially driving it into terrain or other aircraft.

The radio services currently used in civil aviation were simply not designed with malicious parties in mind.  Regulatory agencies and the aviation industry are working together now to develop digital radio protocols for civil aviation, and they haven't done any real work toward security. If the new generation of avionics were widely deployed without cryptographic protection, disaster could result, and the costs of another industry-wide equipment upgrade would be staggering.

We do not suggest that exploitation of navigation vulnerabilities is inevitable or easy. Aircraft are flown by intelligent beings; radio systems are just one element in pilots' and controllers' decision-making processes. There is enough redundancy in the current navigation system that an effective navigation exploit should require multiple simultaneous transmitters. Professional pilots of functional aircraft do not unintentionally hit things that they can see out the window - these problems are specific to flight in instrument meteorological conditions (clouds, fog, precipitation, and smoke).  Still, an attack or threat against airborne radio services would have economic consequences regardless of whether it caused actual physical damage. Commercial aviation only works when customers, employees, and investors all feel safe. Most airline revenue comes from the large, complex terminal areas that are most dependent on communication and navigation; operations would need to be scaled back in the absence of trustworthy radio protocols.

The need for navigation security is not exclusive to aviation. We emphasize examples from that industry because aircraft in flight have enough kinetic and potential energy to cause massive destruction. Surface navigation shares the same essential threats. Also, as explained further below, navigation is largely equivalent to precise time transfer. Time and frequency synchronization over radio is common in communication and surveillance networks, and it is easy to envision time-dependent industrial processes that would go catastrophically wrong if their time scales were distorted or their components were desynchronized.

Improving navigation systems' resilience would have at least one fortuitous positive effect: easing deployment of ultra-wideband waveforms.  Concerns about interference with safety-of-life navigation radio are currently a hindrance to ultra-wideband technology's broad adoption.

\section{Threat environment}

\subsection{Active over-the-air threats}

\begin{description}
\item[Interference] is the accidental masking of signals by other signals or by natural radio emissions. Responding to interference is relatively straightforward and well-studied: multi-transmitter services require protocols for arbitrating multiple access; continuous noise can be countered by directional antennae, spread-spectrum techniques, and increased transmitter power; intermittent noise necessitates error-correction coding or otherwise redundant bitstreams.
\item[Jamming] is the intentional masking of a signal by another transmission, degrading or denying some radio channel. Anti-jam techniques are mostly the same as methods to abate interference, with two additional requirements. First, predicting spread-spectrum variables must be computationally infeasible even given knowledge of all of the transmissions to date or jammers will be able to generate the same sequence and transmit exactly over the content of the signal.  Second, a jammer must not be able to detect the current hop frequency quickly enough to retune to it and transmit before the information-bearing part of the hop's waveform has been entirely sent.\footnote{``Follow-on jamming''.}
\item[Intrusion] is the intentional transmission of false information into a communication or positioning network to deceive or computationally overwhelm targets. ({\bf Spoofing} means roughly the same thing in the context of navigation; it has different connotations in computer networking.) Cryptographic authentication can, by definition, prevent access-granting\footnote{As opposed to denial-of-service.} intrusion on digital radio networks.
\item[Meaconing] is the reception and rebroadcast of legitimate navigation/time signals.\footnote{``Meaconing'' is sometimes used more broadly to mean any sort of intrusion into navigation channels, but a narrow definition is more useful in our context.} Spatially redirecting or precisely delaying signals in a navigation system can lead a victim receiver to an incorrect indication of position. Since meaconing doesn't require attackers to predict signal content, it can't be prevented by merely authenticating the navigation bitstream. Responses to meaconing are the central open problem addressed here.
\end{description}

\subsubsection{Security relationships}

There are relationships between notions of security against these attacks:
\begin{itemize}
\item
      	Any protocol that resists jamming is also robust against interference from benign external sources, because transmitting any signal independent of the target waveform is a subset of assumed jammer capabilities.
\item
        Analogously, an authentication algorithm that prevents intentional intrusion from constituting valid messages also reduces interference and jamming to denial-of-service threats.
\item
        A system that perfectly resists jamming is also immune to intrusion and meaconing: anti-jam security means that a jammer can't transmit {\em anything} to the system's receivers, let alone chosen messages. However, secure communication in the real world still requires authentication, because jamming resistance is conditioned on signal-to-jamming ratio and there is a margin in which a pseudorandom subset of received symbols will be attacker-derived.
\end{itemize}

\subsubsection{Potential source sites for malicious radio transmissions}

\begin{itemize}
\item {\bf On target aircraft}

Ground-based and satellite navaids\footnote{Navigational aids: devices that transmit or reflect a signal to indicate a client's position.} are hundreds to millions of meters away from aircraft radio receivers; passengers and cargo are just a few meters away. Thus, transmitters inside their target aircraft only need thousandths to trillionths of the output power of legitimate navaids to overwhelm them. Aircraft bodies and contents provide some shielding from internal signal sources, but not enough.

Some intrusion and meaconing attacks necessitate precise knowledge of their targets' locations. With transmitters located on the target itself, that becomes vastly easier.

There is no way for typical airport security screeners to distinguish between, for instance, an ordinary laptop and one containing a concealed jammer/spoofer. Portable transmitters can overpower most non-secure airborne radio systems in the host plane (and probably others nearby) for as long as a power source is available.

\item {\bf On other aircraft}

Airborne jammers have some advantages for the attacker over ground-based transmitters: increased exposure of low-altitude targets, increased distance to horizon, and decreased susceptibility to directional antenna nulling.  

Powered craft controlled by malicious parties are easy to locate; in rural areas they are also easy to destroy, but targeting one in an urban environment carries a high risk of civilian casualties.  The airborne threat is not limited to engine-powered vehicles: GPS jamming devices with considerable range that fly on small helium balloons and toy gliders have been demonstrated.\cite{Volpe}  Weather balloons and balloon-launched gliders can rise well above the ceilings of most fighter aircraft. Although above-cloud altitudes leave them visually detectable, they have such a low thermal signature and radar cross-section (unless specifically designed to be radar-reflective) that they're effectively invisible to most seeking technologies. The best available conventional response is anti-radiation missiles (ARMs), which are expensive and limited in availability. Any security force considering an ARM needs to be very sure that the radiating target is not a passenger aircraft. ARMs are not even likely to be effective when wind velocity is relatively high if the attacker uses a constellation of balloon-based jammers that are each only active a fraction of the time.

\item {\bf On the ground}

Ground- and sea-based transmitters have by far the most power available to them\textemdash even a mobile station can output thousands of watts of signal\textemdash and they are capable of disabling all unsecured airborne radio for everything above the horizon.
\end{itemize}

\subsection{Passive over-the-air threats}

The passive threats of concern in military transmission security\footnote{Methods for preventing an adversary from reading from or writing to a communication channel; this primarily refers to spread-spectrum techniques.}\textemdash {\bf detection}, {\bf location}, {\bf identification}, and {\bf interception}\textemdash are of much less importance in civil aviation. Civilian flights operate mostly on schedule between known airports, carrying untrusted passengers and cargo, with no infrared or radar countermeasures and little ability to maneuver to avoid projectiles. One of our major goals in preventing midair collisions is to ensure that aircraft know where all other (untrusted) nearby aircraft are, which obviously negates stealth. Still, we will cursorily examine observability and try to minimize it when possible.

\subsection{Physical threats}

In addition to the threats posed by independently sited malicious transceivers, aviation radio facilities (particularly unstaffed remote navaids) are vulnerable to access-granting physical attacks that subvert cryptologic security:
\begin{itemize}
\item {\bf Navaid movement.} Whether communication channels from a navaid are secure is irrelevant if the navaid can be picked up and relocated.
\item {\bf Directional antenna alteration.} Likewise, navaids that indicate a client's current azimuth or elevation angle can be rotated, be reconfigured, or have their direction sensors changed.
\item {\bf Navaid antenna remoting.} If the transmitter itself is secured or just difficult to move, its transmit or receive antennae could be connected to a long cable and relocated. This isn't a likely risk for high-power transmitters or complex directional antenna arrays, but it is certainly a concern for time-of-flight-based navaids and landing path indicators.
\item {\bf Output meaconing via antenna connection.} Dropping a delay device directly between the navaid processor and its antenna obviates the need to overpower or jam the navaid's original transmission.
\item {\bf Input meaconing via antenna connection.} Navaids that get information about their positions and/or the time via other navaids could be subject to meaconing by devices connected to their antenna inputs; that would be vastly easier than over-the-air meaconing against moving targets.
\item {\bf Deceptive parameter modification.} If the operator interface of a navaid is insufficiently secure, an intruder could change navigation-critical parameters. In timestamped protocols, if the navaid's clock is subject to tampering, much of the protection of cryptographically-secured navigation protocols is negated.
\item {\bf Cryptographic equipment theft.} The entire navaid doesn't necessarily need to be moved\textemdash if cryptographic modules can be covertly stolen and connected to a different transmitter without zeroizing them, the navaid is compromised.
\item {\bf Key compromise.} Navaid keys can be leaked locally through probing, compromising emanations, fault injection, power/timing/heat analysis, and so on.
\end{itemize}

\section{Methods of radionavigation}

A radio signal can be completely characterized by amplitude as a function of position, direction, polarization, and time. For conventional (sinusoidal, non-UWB) transmission systems, it is useful to further decompose time variation into carrier frequency, carrier phase, and modulated content. We can then categorize navigation schemes according to which values are measured by the receiver.

Each of these measurements establishes that the receiver's position is near some geometrical object: a circle if we're measuring distance, a hyperbola for distance differences, a ray or a set of rays for direction.  Actual 3D position is given by the intersection of several similar or different position objects, or sometimes the intersection of position loci with physical objects like the earth's surface.

\subsubsection{Design limitations}

If every client had a continuous channel with every navaid, protected by effective transmission security, we could assume that it was not subject to external manipulation; navigation using such channels would be secure. This security model is appropriate for and currently used by military organizations. However, in a true public service, the combination of thousands of potential users and the large bandwidths consumed by spread-spectrum links would necessitate massively parallel navaid transceivers and huge swaths of spectrum. We need to examine what can be done to assure secure navigation without full transmission security. 

Cost limitations preclude building navigation protocols for which clients require complex directional antenna clusters (phased arrays, etc.). Although that may change as adaptive antenna technology begins to penetrate the consumer market, we will assume simple (omnidirectional or fixed unidirectional) antenna designs for client nodes.\footnote{Building ground transponders for pre-existing airborne traffic and weather radar antennae might be worth considering, though.}

\subsubsection{Cryptographic requirements}

The security protocols that follow presume the existence of a few non-malleable\footnote{Detailed analysis of security requirements on these algorithms is beyond our scope.} cryptographic primitives applicable to messages of arbitrary length:

\begin{itemize}

\item a hash function $H$

\item an integrity-preserving symmetric encryption transform family $e_k, d_k$

\item a public-key signature transform family $S_{priv}, V_{pub}, M: Priv \rightarrow Pub$

\item an integrity-preserving public-key encryption transform family $E_{pub}, D_{priv}, M: Priv \rightarrow Pub$

\end{itemize}

Some sort of certification and revocation architecture is also necessary to bind navaid identifications to public keys, and to let protocol participants determine what degree of confidence in each binding is appropriate.  Any good implementation will provide a flexible language for users and administrators to assign trust in keys, algorithms, equipment certifications, and so on.  We discuss specific certificate contents in section 6, below.

For clarity, we describe some protocols that don't provide anonymity or secrecy.  All of them are readily adaptable by adding an outer layer of encryption or transmission security.

\pagebreak[3]

\vspace{1.4em}
\hrule
\vspace{.2em}
{\Large Passive-client systems}
\vspace{.2em}
\hrule

\subsection{Carrier amplitude}

{\bf Description}
\nopagebreak[3]

In this simplest navigation scheme, navaids transmit a continuous signal.  The received signal strength determines distance to the transmitter.

{\bf Examples}
\nopagebreak[3]

Primitive homing receivers

{\bf Security}
\nopagebreak[3]

Unsecurable. Simply transmitting an unmodulated carrier or a noise signal will change the perceived distance to the navaid, even if the authorized signal is cryptographically authenticable.  In fact, this method is so sensitive to equipment vagaries and environmental propagation conditions that it's not very useful at all. With frequency hopping, the best possible output recovered relative to a jamming environment would be a distance that jumps around the approximate range to the legitimate source as the channel frequency hits other transmitters.

\subsection{Carrier amplitude anisotropy}

{\bf Description}
\nopagebreak[3]

Navaids transmit a simple continuous signal. Receivers use directional antennae to determine the headings to transmitters.

{\bf Examples}
\nopagebreak[3]

NDBs/ADF; S\&R and SIGINT DF receivers 

{\bf Security}
\nopagebreak[3]

Unsecurable in the navaid context. Cost-effective implementations of this technique use simple loop/sense antenna pairs that can't distinguish between multiple emitters. Broadcasting an unmodulated carrier or noise on the beacon frequency will deflect direction measurements made on such an antenna pair.\cite{NDB} Frequency hopping would also be unlikely to help here.

\subsection{Carrier frequency}

{\bf Description}
\nopagebreak[3]

Navaids transmit at a precise known frequency. By measuring actual received frequencies, the client obtains the Doppler shift of each signal and, therefore, the client's velocity relative to each navaid, which determines its position. For this to work, either the client or the navaid needs to be moving, and one or the other must have a predictable position or path.

{\bf Examples}
\nopagebreak[3]

NNSS/Parus, SARSAT/COSPAS, Argos; S\&R and SIGINT DF receivers

{\bf Security}

Unsecurable. Frequency measurements are quite sensitive to narrow-band jamming and intrusion, and again could be altered by transmitting unmodulated signals. Frequency hopping seems unlikely to be compatible with precise frequency measurement of a transmitter not otherwise synchronized with the client, due to oscillator construction limitations.

\subsection{Modulated signal content}

{\bf Description}
\nopagebreak[3]

Navaids transmit anisotropic signals; the content of the received message determines the client's azimuth or elevation angle from the navaid.

Antenna direction can be fixed (as in an indicator that an aircraft is above or below the desired glide path for landing) or varying (as in some general methods for determining azimuth from a navaid). If varying, the directionality can be created by one antenna physically moving in a circle, or an antenna array electronically scanned (circularly or randomly).

{\bf Examples}
\nopagebreak[3]

VOR/Localizer, TACAN azimuth guidance, ILS glide-slope indicator, ILS marker beacons

{\bf Security}
\nopagebreak[3]

Somewhat securable.  The most serious threats to this method are navaid movement, antenna alteration, antenna remoting, and meaconing. Clearly this implies navaids with strong physical security - {\em no unstaffed terrestrial implementation of it can ever be secure}.

Signal-content-based navigation is eminently compatible with all types of transmission security, which would make over-the-air attacks difficult.

The following protocol prevents most easy attacks against systems that can't use transmission security, and adds a degree of resilience to those that can:

\subsubsection{Protocol 1: Passive signal-content navigation} 

\noindent \begin{tabular*}{\columnwidth}{|l|l @{\extracolsep{\fill}} l|}
\hline
\multicolumn{3}{|l|}{\bf Navaid sends:} \\
\hline
$p$ & Navaid position & \\
$d$ & Navaid antenna direction & \\
$t$ & Timestamp & \\
$i$ & Navaid identifier\footnote{An arbitrary value that nodes can use to get public keys from a lookup table.} & \\
$S_k(p, d, t, i)$ & Signature of navigation data & \\
\hline
\end{tabular*}
\vspace{0.1em}

Because a valid signature is not predictable in a reasonable computation time without knowing the navaid's private key $k$, reception of a correct message by the client guarantees three things:

\begin{itemize}
\item At or after the time $t$, a communication channel existed from the indicated navaid's antenna (which the navaid believed to be pointed in the indicated direction at that time) to the client.
\item The current time is at or after $t$ plus the signal's travel time to the aircraft.  So given a time source accurate to the duration of the time step between direction messages (a few milliseconds), a client can exclude older replayed direction packets.\footnote{Note that the timeliness guarantee places an upper bound on the computational workfactor available to the meaconer, so some cryptanalytic and signal-detection techniques are infeasible.}
\item Therefore, a successful meaconer must be able to receive transmissions in realtime from the direction that it wants to tell victims is correct.
\end{itemize}

Meaconing could be done via high-gain antennae that receive sidelobe signals, but that has a partial defense: the navaid can transmit a jamming signal from an omnidirectional antenna co-sited with its directional one. This masking jammer must be significantly stronger at any position available to untrusted equipment than the directional antenna's strongest sidelobe and any information-bearing spurious product or other unintentional emanation of the navigation transmitter. It also must have the same polarization. If site features permit, antennae should be masked to prevent any non-airborne adversary from receiving the directional signal at all.  If possible, aircraft using a runway-mounted navaid for landing guidance should use forward-pointing directional antennae to exclude off-axis attackers.

With these safeguards in place, a node in receipt of a navigation message knows that it was either received directly from the navaid or redirected by a receiver that is airborne and in the antenna's direction right now (though the meaconing transmitter might not be).  Particularly, it should be quite difficult to maliciously tell users of an instrument landing scheme that they're on the correct heading and glide slope to land - an attacker who can insert physical objects into or on an extension of the landing path would be better off just going after the aircraft directly.  As for general azimuth-indicating navaids like VOR, there's little we can do to prevent their signals from being redirected; the passive signal-content technique has limited security unless transmission security is used.

\subsection{Modulated signal timing}

{\bf Description}
\nopagebreak[3]

Navaids either individually transmit a signal indicating the precise current time or cooperate to transmit a simple signal in a precisely-timed sequence. Clients, measuring message arrival times, determine position to be either on a sphere based on absolute time of flight or on a hyperboloid based on difference in time of flight. Time-of-flight-based navigation does not require either navaids or clients to use directional antennae, so it is suitable for mobile peer-to-peer use.

{\bf Examples}
\nopagebreak[3]

GPS, LORAN-C, OMEGA/ALPHA; Link-16/ TADIL-J/MIDS/JTIDS, EPLRS/SADL

{\bf Security}
\nopagebreak[3]

Securable.  The level of assurance is dependent on the number of navaids receivable at the client position.

For small, mutually-trusting node groups, transmission security is readily added to this navigation scheme - some existing protocols already use frequency hopping, time hopping, and direct-sequence spread spectrum together with signal-timing-based navigation.

\subsubsection{Protocol 2: Passive time-of-flight navigation (post-authenticated)}

\noindent \begin{tabular*}{\columnwidth}{|l|l @{\extracolsep{\fill}} l|}
\hline
\multicolumn{3}{|l|}{\bf Navaid sends:} \\
\hline
$p$ & Navaid position & \\
$d$ & Navaid antenna direction & \\
$t$ & Timestamp & \\
$i$ & Navaid identifier & \\
$S_k(p, d, t, i)$ & Signature of navigation data & \\
\hline
\end{tabular*}
\vspace{0.1em}

{\em (Note that navaid transmissions in this protocol are identical to Protocol 1. We examine combining them in a later section.)}

The difference between $t$ and the client's clock value when the client received $t$ is the sum of clock error, propagation time, and meaconing delay.  A client without {\em a priori} knowledge of two of these values can't determine the other one, so ranging or time synchronization based on a single navaid is inherently insecure.  However, each received $t$ establishes a lower bound for the actual current time, and messages from different navaids arrive with little enough time between them that the client can accurately measure the {\em elapsed} time.  If any pair of messages is received with timestamps that indicate the difference between the two is farther than the maximum difference in signal travel time, at least one navaid is being meaconed.  Hence, receiving a single non-meaconed navaid signal sets an upper bound for undetected meaconing delay; unfortunately, that bound represents a distance error on the order of the distance between navaids, so we need to add additional security checks.

Consider a client navigating by multilateration using Protocol 2 alone.  Receiving three navaids specifies a 3D position; receiving four corrects for client clock inaccuracy.  If we add a fifth, we can solve for another variable: the delay time of a single meaconed navaid.  Furthermore, the client can model the measured geometry of the navaids in view and compare it to the position relationships established by the navaids' signed positions; meaconing will make them inconsistent.

Without an independent and highly accurate clock on the client, it is fundamentally impossible to detect using passive techniques alone whether {\em all} of the navaid signals arriving at the client's position are being simultaneously delayed.  It would be technically challenging to remotely specify an arbitrary position to a moving platform, but two replay attacks are easy:
\begin{itemize}
\item placing a simple meaconing device (wideband analog-digital converter, digital delay line, wideband digital-analog converter) on the target platform so that, for instance, it thinks it's at the position in the landing approach that it was a few seconds ago.
\item using a bent-pipe transponder to retransmit the signals received by one client so that others think they have its position. (Note that this is only likely to be persuasive if both clients are airborne or both are ground-based.)
\end{itemize}
These attacks are sufficiently damning in the context of commercial air navigation that passive navigation schemes without transmission security such as GPS should be absolutely ruled out as sole means of navigation for precision approach in zero-visibility conditions.

A differential overlay without transmission security on a passive system adds little or no security, because it's equally susceptible to simple meaconing.  However, if the differential signal is LPI-secure (as in some of the proposed military differential GPS solutions), eliminating some meaconing scenarios is possible.  The security of a combination of mostly untrusted signals with one or two trusted differential signals is an interesting question for further study.

\subsubsection{Protocol 3: Passive time-of-flight navigation (pre-authenticated)}

For vehicles that are so fast and agile that a public-key signature verification delay is unacceptable, it's possible to authenticate timestamps ahead of time:

The navaid generates a random value $r$ and chooses times ($t_1, t_2)$ when it will transmit the following packets.  (We assume that the navaid has previously asserted its identity and position.)

\noindent \begin{tabular*}{\columnwidth}{|l|l @{\extracolsep{\fill}} l|}
\hline
\multicolumn{3}{|l|}{\bf Navaid sends: (Packet 1) (Sent at time $t_1$)} \\
\hline
$H(r, t_2)$ & Hash of $r$ and $t_2$ & \\
$t_1$ & Timestamp of this packet & \\
$S_k(H(r, t_2), t_1)$ & Signature of hash and $t_1$ & \\
\hline
\end{tabular*}
\vspace{0.1em}

The client verifies the signature, stores $H(R, t_2)$, and waits to receive a packet that hashes to that hash.

\noindent \begin{tabular*}{\columnwidth}{|l|l @{\extracolsep{\fill}} l|}
\hline
\multicolumn{3}{|l|}{\bf Navaid sends: (Packet 2) (Sent at time $t_2$)} \\
\hline
$r$ & Random string $r$ & \\
$t_2$ & Timestamp of this packet & \\
\hline
\end{tabular*}
\vspace{0.1em}

The inversion resistance of $H$ guarantees that the node that sent packet 1 knew $r$ and $t_2$; the navaid's signature on $H(r, t_2)$ asserts that the navaid generated $r$ and $t_2$; the unpredictability and novelty of $r$ ensure that no node could have known $r$ before the navaid sent it.

\vspace{1.4em}
\hrule
\nopagebreak[4]
\vspace{.2em}
\nopagebreak[4]
{\Large Active-client systems}
\nopagebreak[4]
\vspace{.2em}
\nopagebreak[4]
\hrule

\subsection{Carrier frequency shift}

{\bf Description}
\nopagebreak[3]

A client transmits a signal on a precisely-known frequency; navaids receive it and respond exactly at the Doppler-shifted frequency they receive, or at some other precisely-related frequency; the client measures the exact received frequency.  Substituting the transmitted and received frequencies into the Doppler equation gives the client's velocity relative to each navaid; that, in turn, can be used to find the client's position.  As in the related client-passive scheme, one or both of the protocol participants must be moving, and one or both must have a predictable position or path.  ``Navaids'' here can be passive\textemdash the most common application of this technique to navigation uses the ground itself as a reflector to determine an aircraft's groundspeed for dead reckoning.

{\bf Examples}
\nopagebreak[3]

Doppler navigation radar; target velocity detection on other radars

{\bf Security}
\nopagebreak[3]

If the interrogation signal is genuinely random, or cryptographically pseudorandom and not repeated, an attacker can't simply transmit a false return signal.  This method is therefore quite secure from meaconing when the ground is used as a reference and the ranging beams are narrow\textemdash an attacker would need to follow the aircraft precisely to supply more than a few seconds' worth of incorrect velocities.  It is significantly less secure if applied to ground-based navaids, where the same attacks that work on the passive version apply.  It might be securable through spread-spectrum modulation, but that would impose strict requirements on oscillator, mixer, and amplifier design.

\subsection{Modulated signal response time}

{\bf Description}
\nopagebreak[3]

A client transmits a signal pattern; navaids return a signal pattern as soon as possible. Subtracting processing delay and multiplying by the wave speed gives the distance between the two transmitters. Navaids for this method can also be passive - for instance, marine radar reflectors and the ground features used by terrain-matching radar.

{\bf Examples}
\nopagebreak[3]

DME, various types of radar

{\bf Security}
\nopagebreak[3]

 Very securable. Both interrogation and response messages can be authenticated, thus making meaconing the only significant vulnerability; the meaconing threat is covered in detail below. As this method is not client-passive, it needs transmission security if low observability is an objective. Frequency hopping and direct-sequence spreading are among the transmission-security methods in use in radar equipment now.

In interactive protocols, the signature and verification delay mentioned in Protocol 3 above is at least doubled, so decoupling the key exchange from the timing-critical segment is actually crucial.

\subsubsection{Protocol 4: Active time-of-flight navigation (pre-authenticated)}

We're designing a public service, so we need to use asymmetric cryptography. Time-based navigation protocols need to happen literally at the speed of light, but asymmetric algorithms are decidedly less quick, especially considering that to avoid timing attacks on private keys we need to fix the time for each operation at its worst-case value. Since we can subtract fixed protocol-induced delays from the message timing used to measure distance, delay error is a matter of how far the platforms can move during the delay rather than how far signals can travel.

Assume two nodes with no prior knowledge of their relative position and velocity are moving together at $10^3$ meters per second.\footnote{This is at least an order of magnitude slower than the fastest airborne objects, but we assume that faster vehicles can accept proportionally higher position uncertainty.}  We want no more than 1 meter of positioning error, so 1 millisecond is the maximum message verification time.  Running two verifications and a signature for a reasonably-secure digital signature algorithm takes several milliseconds on modern general-purpose microprocessors.

However, microsecond-level timing is only crucial within the actual message exchange. It's entirely sufficient for participants to know not where they are right now, but where they were a few milliseconds ago. We can decouple the long-term secure digital signature from the timed message via the following protocol:

The protocol participants agree on a key and symmetric encryption algorithm, and authenticate each other.  Each participant generates a random bit sequence.  The rest of the messages in this protocol are encrypted with the agreed-upon symmetric algorithm.\footnote{If all of the messages in the protocol are quite short, a block cipher in ECB mode is sufficient; otherwise, they'll need to use a robust symmetric encryption mode.}

\noindent \begin{tabular*}{\columnwidth}{|l|l @{\extracolsep{\fill}}l|}
\hline
\multicolumn{3}{|l|}{\bf Client sends: (Interrogation)} \\
\hline
$r_c$ & Client's random string & \\
$t$ & Timestamp & \\
\hline
\end{tabular*}
\vspace{0.1em}

The navaid decrypts each incoming packet with each of the key/cipher pairs that are valid in its area.  Any message that some valid key doesn't decrypt to a valid interrogation or reply is dropped.  If the timestamp is current and the navaid hasn't received that $r_c$ before, it immediately responds:

\noindent \begin{tabular*}{\columnwidth}{|l|l @{\extracolsep{\fill}} l|}
\hline
\multicolumn{3}{|l|}{\bf Navaid sends: (Response)} \\
\hline
$r_c$ & Client's random string & \\
$r_n$ & Navaid's random string & \\
\hline
\end{tabular*}
\vspace{0.1em}

The client measures the precise time from the beginning of its transmission to the end of the navaid's response.  For each valid decryption, it subtracts processing delay to get the round-trip signal time of flight and, therefore, the distance to the navaid.

The navaid's retransmission of the client's newly-generated, unpredictable $r_c$ demonstrates that the navaid received the client's transmission before the client received the navaid's response.  Thus, no meaconer can claim that the difference between the client and navaid is less than it actually is.  All-station meaconing will not work against this protocol as long as one navaid above the number necessary to fix 3D position is within range.

Note that for many key applications\textemdash landing guidance, mid-air traffic avoidance, radio altimetry, and so on\textemdash an attacker being able to increase the measured distance is a {\em critical safety problem}.  A collision-avoidance protocol must therefore either require at least three non-collinear nodes or use transmission security.

Further analysis of the security of similar distance-bounding protocols can be found in \cite{SVLC} and \cite{DBP}.

\subsubsection{Protocol 5: Active time-of-flight navigation (post-autenticated)}

It might be useful in some situations for nodes to determine distance without pre-arranging keys.  In essence, we run the previous protocol without encryption, then authenticate the response afterwards.  This loses authentication of interrogations; that could be restored with the pre-authentication method from Protocol 3, but then this protocol loses any advantage it might have over its pre-authenticated counterpart.

\noindent \begin{tabular*}{\columnwidth}{|l|l @{\extracolsep{\fill}}l|}
\hline
\multicolumn{3}{|l|}{\bf Client sends: (Interrogation)} \\
\hline
$r_c$ & Client's random string & \\
$t$ & Timestamp & \\
\hline
\end{tabular*}
\vspace{0.1em}

\noindent \begin{tabular*}{\columnwidth}{|l|l @{\extracolsep{\fill}} l|}
\hline
\multicolumn{3}{|l|}{\bf Navaid sends: (Response)} \\
\hline
$r_c$ & Client's random string & \\
$r_n$ & Navaid's random string & \\
\hline
\end{tabular*}
\vspace{0.1em}

Then the navaid sends the client a signed, public-key-encrypted message containing $r_c$, $r_n$, its identity, and everything it knew about its position and radiation pattern when it sent its response:

\noindent \begin{tabular*}{\columnwidth}{|l|p{13em} @{\extracolsep{\fill}} l|}
\hline
\multicolumn{3}{|l|}{\bf Navaid sends: (Authentication)} \\
\hline
$p$ & Navaid position at time $t$ & \\
$d$ & Navaid antenna direction at $t$ & \\
$t$ & Interrogation timestamp & \\
$i$ & Navaid identifier & \\
$r_c$ & Client's random string & \\
$r_n$ & Navaid's random string & \\
$S_k(...)$ & Signature of the above &\\
\hline
\end{tabular*}
\vspace{0.1em}

Since third parties can't influence or predict $r_c$ or $r_n$, the signed message demonstrates that the navaid generated its response after it received the interrogation.

\subsection{Modulated signal content}

{\bf Description}
\nopagebreak[3]

A navaid transmits an interrogation signal; the client responds immediately with either a reflection of the interrogation or some other data signal; the navaid then informs the client of its distance and/or azimuth/elevation angle through data or voice messages.

{\bf Examples}
\nopagebreak[3]

Ground-controlled approach

{\bf Security}
\nopagebreak[3]

Securely locating clients works the same and has essentially the same security considerations as clients locating themselves; the remainder of the problem is just transmitting timestamped position messages. That can be done with any type of authentication, encryption, and transmission security.

\subsection{Other methods}

We omit discussion of acoustic and optical means, all of which except celestial navigation are analogues of the radionavigation schemes treated here. There are probably other actual or potential radionavigation methods. Many properties of radio signals are not usable for navigation: Polarization is unsuitable because it is altered by signals' interactions with media and the superposition of multipath signals is not resolvable. Anisotropy in carrier frequency/phase is subject to multipath distortion and not readily measurable with simple antennae. Carrier phase alone only specifies distance inside a wavelength, leaving an integer ambiguity\textemdash though it is useful for refining an existing rough distance measurement; survey-grade GPS receivers use it that way.

\section{Combining navigation techniques}

Navigation is always a matter of combining measurements, and more authenticated measurements means more confidence about position and time.  There are several ways to combine different measurement types that increase security:

\begin{itemize}
\item The navaid transmissions in our first two passive protocols are identical.  A client using Protocol 1 (passive direction-based navigation) can use the timestamps given in that protocol as another Protocol 2 time source, thus adding a level of meaconing resistance to both protocols.

\item Active timing-based ranging by a client and navaid can be daisy-chained into three messages. Each node calculates the range to the other and sends a signed, encrypted measurement of that range; if the ranges are different by more than the nodes could have moved during the delay, imperfect meaconing is strongly indicated.

\item The active timing and angle protocols can also be combined: either a client or a navaid sends out an interrogation, the other one responds immediately, and in the certificated exchange afterward the navaid informs the client of the direction that the client sent its message from.

\item A cleartext passive navigation system such as GPS can be combined with a small number of active-response or passive LPI-secure differential positioning nodes to markedly improve security against meaconing.
\end{itemize}

\section{Non-navigation protocols essential to aviation}

\subsection{Controller-pilot communication links}

The most important factor in preventing mid-air collisions today is voice communication between air traffic controllers and pilots.  Separation procedure and radar surveillance are useless without a channel for instructions, advisories, and questions.

Intrusion on controller-pilot channels (probably in the form of false instructions to pilots) has possibly dire consequences if pilots have no other way to avoid collision.  Imagine two aircraft approaching the same horizontal point with adequate vertical separation; an attacker jams the collision-avoidance channel, orders one of the aircraft to descend to the other's altitude, and immediately jams the controller-pilot connection.

One inventive proposed solution\cite{AIT} implements aircraft authentication and data transfer as a steganographic watermark imposed on the analog voice signal by airborne communication headsets.  However, it appears to be insecure, and it fails to treat the crucial problem of ensuring that a transmission is from an air traffic control facility and reasonably current.  Any secure voice protocol will need to use a digitized voice signal, strongly bind the voice content to a timestamped signature, and provide a key-agreement method.  These are well-solved problems.

Simple jamming is still a serious vulnerability, though, because current procedures for coping with two-way radio communications failure\cite{CFR} rely on leading other traffic out of the unresponsive aircraft's path and landing that aircraft as soon as practicable.  Thus, if aircraft avoided colliding, they would still be expected to terminate their flights at the nearest appropriate airport that they came into visual contact with, which would have an enormous economic impact on commercial airlines.  Some form of anti-jam protection is certainly called for.

One possible solution for non-aerobatic aviation, other than a dedicated secure spread-spectrum voice band, is establishing a text and geospatial-data messaging system that piggybacks on as many other aircraft communication and navigation systems as possible, distributing low-bandwidth messages to every node assumed to be within line of sight of the destination.  An aircraft in receipt of a current authentic packet addressed to its area could also then attempt to forward the packet on a peer-to-peer basis.  Messages would include specific instructions to aircraft, general traffic advisories, weather and turbulence information, cryptographic certificate revocations, and possibly other keying information.  Messages from aircraft to controllers and other aircraft could also be forwarded, on a strict non-interference basis.

Public-key cryptography is applicable to transmission security.  Each node could have several relatively simple spread-spectrum transceivers and could use a key-agreement protocol to establish transmission security keys with navaids, air traffic control facilities, and peer aircraft via any other available node.  Alternatively, generating transmission security keys could be part of filing a flight plan.  Each landing-guidance installation could have a radio that toggled between the transmission security parameters for expected inbound aircraft, and several radios that communicated with the nearest few craft.

Another simple way to decrease the impact of jammers on communication is equipping each aircraft with two antennae with hemispherical reception patterns, one pointed up and one down.\footnote{TCAS does something like this, and the F-16 has upper and lower antennae for UHF voice and SADL, but neither implementation seems to be designed to adaptively detect and avoid jamming.}  For aircraft with normal en-route attitudes, this prevents jammers above the aircraft (like balloons) from blocking signals from below (like landing-guidance transmitters) and vice versa, and is especially powerful when combined with the above peer-to-peer distribution network in the presence of a mixture of ground-based, satellite, and airborne communication nodes.

\subsection{Identification protocols}

Civilian aircraft are identified by a 4-digit transponder code assigned on the fly by air traffic controllers and entered by pilots.  The cockpit-selectable transponder ID system is a waste of pilot and controller time and a security risk. Aircraft should be identified by a globally unique code at all times; anonymity of general aviation flights can be preserved by encrypting transponder transmissions.  The newer Mode S standard uses globally unique identifiers, but provides no anonymity and no protection against impersonation.

Military aircraft in the USA and USA-allied nations use the Mark XII IFF protocol to identify airborne objects.  Mark XII is (weakly) cryptographically authenticated but has no transmission security; it could easily be denied by jamming. Since lack of IFF response is sometimes taken to indicate a hostile aircraft, IFF jamming in a complex battlefield setting could lead to fratricide. Several friendly-fire incidents attributable in part to IFF malfunctions have happened in the last few years.\cite{CRS-IFF}

\subsection{Position-reporting protocols}

The current paradigm in aviation protocols is ADS-B,\footnote{Automatic Dependent Surveillance - Broadcast} in which aircraft determine their position by some method (generally assumed to be GPS) and then broadcast their identification, location, and various other data.  Traffic controllers and pilots can use the position reports to plan optimal routes while ensuring aircraft separation, which promises to markedly increase efficiency and ease psychological workloads, among a host of other benefits.

Of course this is only as secure and reliable as the positioning sources that it depends on.  As for ADS-B itself, it should be trivially easy to add encrypted versions of the payload packets for clients who require anonymity to peers, a ``key agreement'' packet type to handle cryptographic setup, and a ``digital signature'' packet type that aggregates timestamped signatures for several recent packets.  An additional packet type should be defined for airborne and ground nodes to report discrepancies between GPS position and environmental observables, and between other nodes' stated and radar-measured positions.

A potential concern with frequent automatic position reporting is that it gives attackers a feedback channel into how their meaconing is affecting the targets' notion of position.  Also, the TIS-B protocol, which is part of the ADS-B suite, has ground radar facilities transmittting locations of aircraft as determined from primary and secondary radar and ADS.  While this is wonderful for safety, it also provides malicious users with an oracle into their radar detectability and the efficacy of ECM techniques, so its costs and benefits require careful consideration.

\subsection{Emergency rescue transponders}

Distress-beacon detection satellites (SARSAT/ COSPAS) and rescue crews have limited capacity and no authentication system.  An adversary who wanted to prevent downed aircraft from being located could do so quite easily by activating rogue emergency location transmitters (ELTs) in the general vicinity of the actual distressed craft, potentially with the same identification information as the legitimate ELTs.  Fake ELTs can be built cheaply.

\section{Certificate structure}

Navigation involves many changing security-related values and little spare bandwidth, so the sort of monolithic certificate used in most Internet protocols is not really appropriate.  We suggest the use of a modular certificate format, with small certificates attesting to a small set of facts, transmitted upon request and/or broadcast periodically.  All certificates should contain:

\begin{itemize}
\item A set of assertions
\item The time range in which the assertions are valid
\item The identifier of the node about which assertions are being made
\item The public key of the certifier
\item A signature of a hash of the certificate
\end{itemize}

Some common assertions are:

\begin{itemize}
\item {\em Navigation-related information:} Position, antenna pointing angle (as functions of time - for instance, Keplerian orbital elements for satellites)
\item {\em Identifying information:} Tail number, aircraft type and model
\item {\em Cryptographic delegation information:} Public keys of nodes
\item {\em Information describing why someone should trust the node's public key:} Key-holding device manufacturer/model, cryptographic security type (unverified, remotely secure, tamper-resistant)\cite{FIPS140}, cryptographic verification authority, physical security level (unsecured, alarmed, sealed, supervised), owner, owner type (individual, airline, etc.)
\item {\em Information that can help with non-cryptographic security verification:} \cite{WJ},\cite{WPK04} Platform type (surface, airborne, etc.), receiver sensitivity and transmitter power output (given as spherical functions of antenna pointing angle)
\end{itemize}

\section{Acknowledgements}

Useful technical and policy advice was provided by Demoz Gebre-Egziabher, Carol Kohtz, David Freeman, and Josh Cohen.

\end{multicols}

\end{document}